\documentclass[prd,twocolumn,superscriptaddress,showpacs]{revtex4}
\usepackage{graphicx}
\usepackage{times}
\usepackage{amsmath}
\usepackage{bm}
\usepackage{amsfonts}
\usepackage{amssymb}
\usepackage{newlfont}
\usepackage{amsthm}

\begin{document}

\title{Dissipation due to fermions in
 inflaton equations of motion}
\author{Javier~Moreno Almeida}
\email[email:~]{javier.moreno@indizen.com}
\author{Ian D.~Lawrie}
\email[email:~]{i.d.lawrie@leeds.ac.uk}

\affiliation{School of Physics \&\ Astronomy, University of Leeds,
Leeds, LS2 9JT UK}

\begin{abstract}
According to quantum field theory, the inflaton equation of motion
does not have the local form that is generally assumed for
cosmological purposes.  In particular, earlier investigations of the
nonequilibrium dynamics of an inflaton that decays into scalar particles
suggest that the loss of inflaton energy is not well approximated
by the local friction term derived from linear response theory.  We extend
this analysis to the case of an inflaton that decays into fermions,
and reach broadly the same conclusion.
\end{abstract}

\date{\today}
\maketitle
\section{introduction}\label{introduction}
In the study of inflationary cosmology, it is often assumed that a
classical inflaton field $\phi$ obeys a local equation of motion
of the form
\begin{equation}\label{localeom}
\ddot{\phi} + \eta(\phi)\dot{\phi}+V'(\phi)=0,
\end{equation}
where the friction term $\eta(\phi)\dot\phi$ accounts for the loss
of inflaton energy by radiation of particles.  (We omit the usual
Hubble damping term $3H\dot\phi$, for reasons discussed below.)
In particular, ``warm inflation'' scenarios (see, for example,
\cite{moss2,berera,bererafang,bereraNP,moss}) envisage a friction
coefficient $\eta(\phi)$ large enough for the inflaton motion
to be perpetually overdamped, thus avoiding the usual process
of reheating.

However, an equation of this form does not arise automatically from
quantum field theory.  Indeed, if the inflaton couples biquadratically
to scalar fields $\chi$, and via Yukawa interactions to spin-$\frac{1}{2}$
fields $\psi$, its equation of motion is generically of the form
\begin{equation}
\ddot{\phi}+V'(\phi)
+g_1\langle\chi^2\rangle\phi+g_2\langle\bar{\psi}\psi\rangle+\cdots
=0 \label{phieom}
\end{equation}
where the expectation values $\langle\chi^2\rangle$ and
$\langle\bar{\psi}\psi\rangle$ are nonlocal functionals of $\phi$.
That is, they depend on the entire history of $\phi$, at all times
prior to the time $t$ at which the expectation values are evaluated.
The issue we address here is whether these expectation values can
be adequately approximated by a local friction term, together with
a local correction to the effective potential $V(\phi)$, leading to
the equation of motion (\ref{localeom}).  To focus on this
issue, which is substantially independent of the background metric,
we simplify matters throughout by studying quantum field theory in
Minkowski spacetime, with $H=0$.

If the motion of $\phi$ is slow enough, it seems intuitively
reasonable that suitable approximations to
these expectation values can be obtained by considering the linear
response of a state of thermal equilibrium to a time-dependent
perturbation, $\dot\phi(t)$.  Several implementations of this
general idea can be found in the literature (see, for example
\cite{moss2,hosoya,morikawa,morikawa2,bereraramosgleiser,bereraramos}), and
lead to well-defined estimates of the friction coefficient $\eta(\phi)$.
However, the {\it assumption} that this problem can adequately
be treated on the basis of linear perturbations about {\it equilibrium}
thermal field theory deserves further investigation.  If this assumption
is valid, then the results should be reproduced in the slow evolution
limit of {\it nonequilibrium} field theory.

For the case of an inflaton
coupled to a scalar field, this does not seem to be true
\cite{lawrie02,lawrie,lawrienumeric}.  Formally, one can derive
approximate, local nonequilibrium evolution equations, but a time-derivative
expansion of the solution to these equations, which is needed to
arrive at (\ref{localeom}), does not exist.  With the further approximation
of replacing non-equilibrium self-energies with their equilibrium
counterparts, a time-derivative expansion becomes possible, leading
to what we call the {\it adiabatic approximation}, which essentially
reproduces the results of linear response theory.  However, numerical
investigation shows that the motion generated by the nonequilibrium equations is
not well approximated by the adiabatic approximation.

These results might well be specific to the particular field theory
considered.  In this paper, we investigate this to some extent by
extending the analysis of \cite{lawrie02,lawrie,lawrienumeric} to the
case of an inflaton that decays into spin-$\frac{1}{2}$ fermions.  In
principle, the calculations are quite similar, but the technicalities
associated with spin-$\frac{1}{2}$ fields are slightly more complicated.

In section \ref{freefer} we derive exact evolution equations for
the number densities $N_k(t)$, and auxiliary functions
$\nu_k(t)$, which together specify the nonequilibrium state of fermions
that are free, except for a $\phi$-dependent mass arising from a
Yukawa coupling to the classical inflaton. These are the fermionic analogues
of the exact equations derived for scalar fields by Morikawa and
Sasaki \cite{morikawa}.  As in the scalar-field case, they do not describe
any significant frictional effect, because particles radiated by
the inflaton motion can easily be reabsorbed.  To achieve a
permanent transfer of energy from the inflaton to the system of
particles, these particles must interact with each other, and dissipate
energy by scattering.

An approximate set of nonequilibrium evolution equations is obtained
in section \ref{interactions}, by adapting the analysis described by
Lawrie and McKernan in \cite{lawriefermion}.  The closed-time-path
methods used to develop this approximation are very different from the
straightforward (if long-winded) operator methods available for free
field theories, and the fact that the equations of section \ref{freefer}
are recovered in the free-field limit is a crucial check on this method
of approximation.  We again find that the nonequilibrium evolution equations
do not admit a solution in the form of a time-derivative expansion
(in fact, this seems to be a general, model-independent result),
but an adiabatic approximation can be devised, as explained in section
\ref{adiabatic}.

Numerical solutions to the nonequilibrium evolution equations, and
to the adiabatic approximation to these equations, are described
in section \ref{results}.  As in the previous case of an inflaton that
decays into scalars, we find that these solutions do not agree well.
In general, the adiabatic approximation appears to overestimate
the damping effect of particle radiation on the inflaton motion;
in particular, we find that overdamping predicted by the
adiabatic approximation, is not exhibited by the nonequilibrium
evolution.

A self-contained summary and discussion of these results is given in section
\ref{conclusions}.

\section{Free-Field Evolution Equations}\label{freefer}
In this section, we derive a set of exact equations for the time evolution of a
system of fermions that are free, apart from a coupling to the
classical inflaton;  this gives rise to an effective time-dependent
mass $m(t)=m_{\psi}+g\phi(t)$, where $m_{\psi}$ denotes the bare mass.
These equations serve as a valuable check on the validity and
interpretation of the approximate evolution equations we obtain in section
\ref{interactions} for the system of interacting fermions.

We look for a formal solution of the Dirac equation with time dependent mass,
\begin{equation}\label{dirac}
\left[i\gamma^{\mu}\partial_{\mu}-m(t)\right]\psi(x)=0
\end{equation}
by expanding the field $\psi(x)\equiv\psi(\bm{x},t)$ in terms of two positive-energy spinors $u_{\pm}$
and their charge conjugates $u^c_{\pm}=i\gamma^2u_{\pm}^*$:
\begin{eqnarray}\label{psiexpansion}
\psi(x)&=&\int\frac{d^3k}{(2\pi)^3}e^{i\bm{k}\cdot\bm{x}}[A_k(t;\hat{t})u_+
(\bm{k},\hat{t})+B_k(t;\hat{t})u_-(\bm{k},\hat{t})\nonumber\\
&&+C_k(t;\hat{t})u_+^c(-\bm{k},\hat{t})+D_k(t;\hat{t})u_-^c(-\bm{k},\hat{t})].
\end{eqnarray}
The spinors $u_{\pm}(\bm{k},\hat{t})$ are helicity eigenstates, with helicity
$\frac{1}{2}h$, $h=\pm 1$, as indicated by the subscript and, at an arbitrary reference time
$\hat t$, are solutions of the instantaneous Dirac equation
\begin{equation}
\left[\gamma^0\hat\Omega_k+\bm{\gamma\cdot k}+\hat{m}\right]u_\pm(\bm{k},\hat{t})=0,
\end{equation}
where $\hat{m}=m(\hat{t})$ and $\hat{\Omega}_k=\sqrt{k^2+\hat{m}^2}$, normalized such
that $u_h^\dag(\bm{k},\hat{t})u_{h'}(\bm{k},\hat{t})=\delta_{h h'}$.
Dependence on the time $t$ is contained in the functions
$A_k(t;\hat{t}), B_k(t;\hat{t}),\ldots$ and these coefficients also depend on $\hat{t}$,
in such a way that $\psi(x)$ itself is independent of $\hat{t}$. By substituting the
\textit{Ansatz} (\ref{psiexpansion}) into the Dirac equation (\ref{dirac}),
we find equations for the $t$ dependence of $A_k(t;\hat{t}),
B_k(t;\hat{t}),\ldots$. The change of variables
\begin{equation}
\begin{aligned}
{A_k}&=c_k\tilde{A}_k-s_k\tilde{C}_k & B_k&=c_k\tilde{B}_k-s_k\tilde{D}_k\\
C_k&=s_k\tilde{A}_k+c_k\tilde{C}_k & D_k&=-s_k\tilde{B}_k-c_k\tilde{D}_k
\end{aligned}
\label{abcdtrans}
\end{equation}
with
\begin{equation}
\begin{aligned}
c_k&=\frac{\hat{\Omega}_k+\hat{m}}{\sqrt{2\hat{\Omega}_k(\hat{\Omega}_k+\hat{m})}} &
s_k&=\frac{\vert\bm{k}\vert}{\sqrt{2\hat{\Omega}_k(\hat{\Omega}_k+\hat{m})}}
\end{aligned}
\end{equation}
(satisfying $c_k^2+s_k^2=1$) brings these equations into a
standard form used in \cite{lawriefermion}, namely
\begin{equation}\label{idt}
\begin{aligned}
i\partial_t
\begin{pmatrix}
\tilde{A}_k(t;\hat{t})\\
\tilde{C}_k(t;\hat{t})
\end{pmatrix}
&=\bm{T}_k(t)
\begin{pmatrix}
\tilde{A}_k(t;\hat{t})\\
\tilde{C}_k(t;\hat{t})
\end{pmatrix}\\
i\partial_t
\begin{pmatrix}
\tilde{B}_k(t;\hat{t})\\
\tilde{D}_k(t;\hat{t})
\end{pmatrix}
&=\bm{T}_k(t)
\begin{pmatrix}
\tilde{B}_k(t;\hat{t})\\
\tilde{D}_k(t;\hat{t})
\end{pmatrix}.
\end{aligned}
\end{equation}
The matrix
\begin{equation}
\bm{T}_k(t)=
\begin{pmatrix}
m(t) & -k\\
-k & -m(t)
\end{pmatrix}
\end{equation}
(with $k=\vert\bm{k}\vert$) has eigenvalues $\pm\Omega_k(t)$, where
$\Omega_k(t)=\sqrt{m(t)^2+k^2}$. Correspondingly, two orthogonal
solutions to the equation $i\partial_t\bm{F}(t)=\bm{T}_k(t)\bm{F}(t)$ can
be written as
\begin{eqnarray}
\bm{F}^{(+)}_k(t;\hat{t})&\equiv&
\begin{pmatrix}
f_k(t;\hat{t})\\
g_k(t;\hat{t})
\end{pmatrix}
=\cal{T}e^{-i\int_{\hat{t}}^t\bm{T}_k(t')dt'}
\begin{pmatrix}
c_k\\
-s_k
\end{pmatrix}\\
\bm{F}^{(-)}_k(t;\hat{t})&\equiv&
\begin{pmatrix}
-g^*_k(t;\hat{t})\\
f^*_k(t;\hat{t})
\end{pmatrix}
=\cal{T} e^{-i\int_{\hat{t}}^t\bm{T}_k(t')dt'}
\begin{pmatrix}
s_k\\
c_k
\end{pmatrix}
\end{eqnarray}
where $\cal{T}$ denotes time ordering. At the reference time $\hat{t}$,
$\bm{F}^{(+)}_k$ and $\bm{F}^{(-)}_k$ are positive- and negative-energy solutions
respectively (that is, $\left.i\partial_t\bm{F}^{(\pm)}_k\right\vert_{t=\hat{t}}
=\pm\hat{\Omega}_k\bm{F}^{(\pm)}_k$).

The general solution
for $\tilde{A}_k(t;\hat{t}), \tilde{B}_k(t;\hat{t}), \ldots$ can thus
be written as
\begin{eqnarray}
\begin{pmatrix}
\tilde{A}(t;\hat{t})\\
\tilde{C}(t;\hat{t})
\end{pmatrix}
&=&b_+(\bm{k},\hat{t})
\begin{pmatrix}
f_k(t;\hat{t})\\
g_k(t;\hat{t})
\end{pmatrix}
+d_+^{\dag}(-\bm{k},\hat{t})
\begin{pmatrix}
-g^*_k(t;\hat{t})\\
f^*_k(t;\hat{t})
\end{pmatrix}\nonumber\\
\begin{pmatrix}
\tilde{B}(t;\hat{t})\\
\tilde{D}(t;\hat{t})
\end{pmatrix}
&=&b_-(\bm{k},\hat{t})
\begin{pmatrix}
f_k(t;\hat{t})\\
g_k(t;\hat{t})
\end{pmatrix}
-d_-^{\dag}(-\bm{k},\hat{t})
\begin{pmatrix}
-g^*_k(t;\hat{t})\\
f^*_k(t;\hat{t})
\end{pmatrix}\nonumber\\
\label{abcdsolution}
\end{eqnarray}
and the canonical equal-time anticommutator of $\psi(x)$ and $\psi^\dag(x)$
implies that the coefficients $b_h(\bm{k},\hat{t})$ and $d_h(\bm{k},\hat{t})$
have the usual anticommutator algebra
$$
\{b_h(\bm{k,\hat{t}}),b_{h'}^{\dag}(\bm{k}',\hat{t})\}=(2\pi)^3\delta(\bm{k}-\bm{k}')\delta_{hh'},
\text{etc}
$$
for particle creation and annihilation operators.

To find how these operators depend on $\hat{t}$ we make use of the fact that
$\partial_{\hat t}\psi(x)=0$.  On substituting (\ref{abcdtrans}) and (\ref{abcdsolution})
into (\ref{psiexpansion}), we find (after some algebra) that
\begin{equation}
\begin{aligned}
\partial_{\hat{t}}b_h(\bm{k},\hat{t})&=-i\hat{\Omega}_kb_h(\bm{k},\hat{t})
+h\lambda_kd_h^\dag(-\bm{k},\hat{t})\\
\partial_{\hat{t}}d_h(\bm{k},\hat{t})&=-i\hat{\Omega}_kd_h(\bm{k},\hat{t})
-h\lambda_kb_h^\dag(-\bm{k},\hat{t})
\end{aligned}\label{dbdthat}
\end{equation}
where
\begin{equation}
\lambda_k=\frac{k\partial_{\hat t}\hat{\Omega}_k}{2\hat{\Omega}_k\hat{m}}
=\frac{k\partial_{\hat t}\hat{m}}{2\hat{\Omega}_k^2}.
\end{equation}
In a translation- and rotation-invariant state, expectation values of bilinear products
of creation and annihilation operators are characterized by functions $N_h(k,\hat{t})$,
$\bar{N}_h(k,\hat{t})$ and $\nu_h(k,\hat{t})$, defined by
\begin{equation}
\begin{aligned}
\langle
 b_h^{\dag}(\bm{k},\hat{t})b_{h'}(\bm{k}',\hat{t})\rangle&=(2\pi)^3\delta(\bm{k}-\bm{k}')\delta_{hh'}N_h(k,\hat{t})\\
\langle
 d_h^{\dag}(\bm{k},\hat{t})d_{h'}(\bm{k}',\hat{t})\rangle&=(2\pi)^3\delta(\bm{k}-\bm{k}')\delta_{hh'}\bar{N}_h(k,\hat{t})\\
\langle
d_h(-\bm{k},\hat{t})b_{h'}(\bm{k}',\hat{t})\rangle&=(2\pi)^3\delta(\bm{k}-\bm{k}')\delta_{hh'}\nu_h(k,\hat{t}).
\end{aligned}\label{densities}
\end{equation}
Because the mode functions $\bm{F}^{(+)}_k(t,\hat{t})$ are positive-frequency only for times $t$
near $\hat{t}$, we can interpret $N_h(k,\hat{t})$ as the number density of particles present
\textit{at time} $\hat{t}$ with momentum $\bm{k}$ and helicity $h$, and $\bar{N}_h(k,\hat{t})$
as the corresponding density of antiparticles.  The functions $\nu_h(k,\hat{t})$ measure the
off-diagonality of the density matrix in the basis specified by the modes $\bm{F}^{(\pm)}_k$,
and have no classical analogue.

The expectation value $\langle\bar{\psi}(\bm{x},t)\psi(\bm{x},t)\rangle$
is conveniently expressed in terms of the number densities at time $t$, and we achieve this by setting
$\hat{t}=t$.  We will restrict attention to states in which
$N_+(k,t), N_-(k,t), \bar{N}_+(k,t)$ and $\bar{N}_-(k,t)$ are all equal, say to $N_k(t)$,
while $\nu_+(k,t)=-\nu_-(k,t)\equiv\nu_k(t)$.
It is easily checked using (\ref{dbdthat}) that these equalities are preserved by the time evolution,
and that the real number density $N_k(t)$ and the complex function $\nu_k(t)$ obey the evolution
equations
\begin{eqnarray}
\partial_tN_k(t)&=&\frac{k\dot{m}(t)}{\Omega_k^2(t)}\nu^R_k(t)\label{dndtferI}\\
\partial_t\nu_k(t)&=&-2i\Omega_k(t)
\nu_k(t)-\frac{k\dot{m}(t)}{2\Omega_k^2(t)}\left[2N_k(t)-1\right],\nonumber\\
\label{dnudtferI}
\end{eqnarray}
where $\nu^R$ is the real part of $\nu$.  The expectation value of interest
is given in terms of these functions by
\begin{equation}\label{psibarpsi}
\langle
\bar{\psi}(\bm{x},t)\psi(\bm{x},t)\rangle
=4\int\frac{d^3k}{(2\pi)^32\Omega_k}\left[m(2N_k-1)+2k\nu_k^R\right],
\end{equation}
with all the quantities $m$, $\Omega_k$, $N_k$, $\nu_k$ evaluated at time $t$.
The evolution equations (\ref{dndtferI}) and (\ref{dnudtferI}) are the fermionic analogues
of those derived by Morikawa and Sasaki \cite{morikawa} for the case of the inflaton
field coupled to a scalar field. The right-hand side of (\ref{dndtferI}) is a
particle `creation' term, proportional to $d\phi/dt$, but it can be either
positive or negative as $\nu^R_k$ oscillates, and a numerical solution shows that
there is no permanent transfer of energy from the inflaton to the system of fermions.  To see
a frictional effect, it is necessary to introduce interactions between the fermions
so that, roughly speaking, the created particles lose coherence with the inflaton,
and a permanent dissipation of energy results.
\section{Evolution Equations for an
interacting system}\label{interactions}
\subsection{Derivation}\label{deriv}
An approximation scheme for studying the non-equilibrium evolution
of a system of interacting fermions was developed by Lawrie and
McKernan in \cite{lawriefermion}. We follow their derivation closely,
except that we introduce, as in section \ref{freefer}, an evolving
reference time $\hat{t}$, which allows us to make explicit the effect
of particle creation.

Within the closed-time-path (CTP) formalism (see, e.g. \cite{semenoff,kobes} and
references therein), we seek an approximation to the full
two-point function
\begin{eqnarray}
&&\cal{S}_{\alpha\beta}^{(ab)}(t,\bm{x};t',\bm{x}')=\nonumber\\
&&\begin{pmatrix} \langle
\mathcal{T}[\psi_{\alpha}(t,\bm{x})\bar{\psi}_{\beta}(t',\bm{x}')]\rangle &
-\langle \bar{\psi}_{\beta}(t',\bm{x}')\psi_{\alpha}(t,\bm{x})\rangle\\
\langle
\psi_{\alpha}(t,\bm{x})\bar{\psi}_{\beta}(t',\bm{x}')\rangle &
\langle
\bar{\mathcal{T}}[\psi_{\alpha}(t,\bm{x})\bar{\psi}_{\beta}(t',\bm{x}'))]\rangle
\end{pmatrix}\nonumber\\
\label{Sdef}
\end{eqnarray}
from which the expectation value $\langle\bar{\psi}(\bm{x},t)\psi(\bm{x},t)\rangle$
can be obtained as a special case. (Here, $\alpha, \beta$ are Dirac indices, while $a, b
=1,2$ distinguish the indicated operator orderings.)
It is shown in \cite{lawriefermion} that this can be expressed (after a
spatial Fourier transform) in terms of
a single $4\times 4$ matrix $\cal{H}(t,t';\bm{k})$ as
\begin{eqnarray}
\cal{S}^{(ab)}(t,t';\bm{k})&=&
\begin{pmatrix}
\cal{H}(t,t';\bm{k}) & \tilde{\bar{\cal{H}}}(t,t';\bm{k})\\
\cal{H}(t,t';\bm{k}) & \tilde{\bar{\cal{H}}}(t,t';\bm{k})
\end{pmatrix}
\theta(t-t') \nonumber\\
&+&\begin{pmatrix}
\tilde{\cal{H}}(t,t';\bm{k}) & \tilde{\cal{H}}(t,t';\bm{k})\\
\bar{\cal{H}}(t,t';\bm{k}) & \bar{\cal{H}}(t,t';\bm{k})
\end{pmatrix}
\theta(t'-t). \label{propfourier}
\end{eqnarray}
For any matrix $M$, we define $\bar{M}=\gamma^0M^{\dag}\gamma^0$ and
$\tilde{M}=[C^{-1}\gamma^0M\gamma^0C]^T$ where $C$ is the charge
conjugation matrix and $^T$ indicates the transpose.

As explained in detail in \cite{lawriefermion} (and in
\cite{lawrie2,lawrie} for the case of $\phi$ coupling to a scalar field),
we construct a partially-resummed perturbation theory as follows.  Given
a CTP action $I=I^{(2)}+I^{(>2)}$, where $I^{(2)}$ is quadratic in the fields
and $I^{(>2)}$ is of higher order, we take its lowest-order part to be
\begin{equation}
I_0(\psi)=I^{(2)}(\psi)+\int
d^4x\bar{\psi}_a\cal{M}_{ab}\psi_b\equiv\int
d^4x\bar{\psi}_a\cal{D}_{ab}\psi_b.\label{i0}
\end{equation}
where $\mathcal{M}$ is a counterterm to be determined, and treat the remainder,
$I_{\mathrm{int}}=I^{(>2)}-\int d^4x\bar{\psi}_a\cal{M}_{ab}\psi_b$ as
a perturbation. (For $a=1,2$, $\psi_a$ are the independent path-integration variables
that inhabit the real-time branches of the closed time path;  we have indicated
explicitly only the contribution of one spin-$\frac{1}{2}$ field.)
The lowest order propagator $S(t,t')$ (we suppress the momentum argument $\bm{k}$)
obeys the equation
\begin{eqnarray}
\cal{D}_{ac}(t,\partial_t)S^{(cb)}(t,t')&=&S^{(ac)}(t,t')\cal{D}_{cb}(t',-\overleftarrow{\partial}_{t'})\nonumber\\
&=&i\delta_{ab}\delta(t-t')\label{diffop}
\end{eqnarray}
where the differential operator $\cal{D}$, defined by (\ref{i0}), is
\begin{equation}\label{doperator}
\cal{D}(t,\partial_t)=
\begin{pmatrix}
\mathcal{D}_0(t,\partial_t) & 0\\
0 & -\mathcal{D}_0(t,\partial_t)
\end{pmatrix}
+\mathcal{M}(t),
\end{equation}
and $\mathcal{D}_0=i\gamma^0\partial_t-\bm{\gamma\cdot k}-m(t)$ is the usual Dirac operator.
Given suitable constraints on the form of the counterterm $\mathcal{M}$,
the solution for $S^{(ab)}(t,t')$ can be written in the form (\ref{propfourier})
in terms of a function $H(t,t')$ which is the lowest-order approximation to
$\cal{H}(t,t')$.

In this way, we obtain a reorganized perturbation theory, in which the
vertices implied by $I^{(>2)}$ are supplemented by the counterterm $\mathcal{M}$.
In particular, self-energies have the form
\begin{equation}
\Sigma_{ab}(t,t')=\mathcal{M}_{ab}(t)\delta(t-t')+\Sigma_{ab}^{\mathrm{loop}}(t,t')
\end{equation}
and we can optimize the propagators $S^{(ab)}$ as an approximation to the full
2-point functions $\mathcal{S}^{(ab)}$ by choosing $\mathcal{M}_{ab}$ to cancel
a local part of the loop diagrams in $\Sigma_{ab}^{\mathrm{loop}}$.

The solution of (\ref{diffop}) is explained in detail in \cite{lawriefermion}.
The matrix $H(t,t')$ from which $S^{(ab)}(t,t')$ is built can be expressed
in terms of the $\gamma$ matrices, the mode functions $\bm{F}_k^{(\pm)}(t,\hat{t})$
introduced in section \ref{freefer}, and two auxiliary functions
$N_k(t,\hat{t})$ and $\Delta_k(t,\hat{t})$ which, in the free-field
limit, reduce to $N_k(\hat{t})$ and $-\nu_k(\hat{t})$ respectively.  These
functions satisfy differential equations in the variable $t$, which arise from
(\ref{diffop}). In \cite{lawriefermion}, the reference time $\hat{t}$ was taken
to be fixed at at the instant when an initial state is specified, but
here we find it convenient to allow $\hat{t}$ to vary, and eventually, as
in section \ref{freefer}, to set $\hat{t}=t$.  The dependence of $N_k(t,\hat{t})$
and $\Delta_k(t,\hat{t})$ on $\hat{t}$ can be found from the fact that
$S^{(ab)}(t,t')$ must be independent of $\hat{t}$.

We now define
\begin{equation}
N_k(t)=N_k(t,t)\qquad\hbox{and}\qquad\nu_k(t)=-\Delta_k(t,t).
\end{equation}
Evidently, we have
\begin{equation}
\partial_tN_k(t)=\left[\partial_tN_k(t,\hat{t})+\partial_{\hat{t}}N_k(t,\hat{t})
\right]_{\hat{t}=t}
\end{equation}
and similarly for $\nu_k(t)$.  Applying the results of \cite{lawriefermion},
we find that these two equations have the form
\begin{eqnarray}
\partial_{t}N_k(t)&=&-\Gamma_k(t)\left[2N_k(t)-1\right]+\alpha_k(t)+ \frac{k\dot{m}(t)}{\Omega_k(t)^2}
\nu_k^R(t)\nonumber\\
\label{dndtfermion}\\
\partial_{t}\nu_k(t)&=&-2[i\Omega_k(t)+\Gamma_k(t)]\nu_k(t)\nonumber\\
&&-\frac{k\dot{m}(t)}{2\Omega_k(t)^2}\left[2N_k(t)-
1\right].\label{dnudtfermion}
\end{eqnarray}
The functions $\Gamma_k(t)$, which can be interpreted as a quasiparticle
decay width, and $\alpha_k(t)$ appear in the counterterm $\mathcal{M}$ and,
according to the strategy outlined above, are to be determined from the
self-energies $\Sigma_{ab}^{\mathrm{loop}}$.

A concrete realization of this strategy is described in \cite{lawriefermion}
for the case that the fermions interact through a Yukawa coupling to a
scalar field $\chi(\bm{x},t)$ of mass $M$,
\begin{equation}
\mathcal{L}_{\psi\chi}=-g_\chi\bar{\psi}(\bm{x},t)\psi(\bm{x},t)\chi(\bm{x},t),
\label{yukawa}
\end{equation}
and we consider the same model here, assuming for simplicity that $\chi$ does not
couple to the inflaton, so $M$ is constant.  Evaluating the relevant self-energies
to 1-loop order gives
\begin{eqnarray}
\Gamma_k(t)&=&\frac{g_\chi^2}{64\pi^2}(M^2-4m^2)\int
d^3k'\frac{\delta(\omega_p-\Omega_k-\Omega_{k'})}{\Omega_k\Omega_{k'}\omega_p}\nonumber\\
&&\times\left[n_p+N_{k'}\right]\label{Gammapsi}\\
\alpha_k(t)&=&\frac{g_\chi^2}{64\pi^2}(M^2-4m^2)\int
d^3k'\frac{\delta(\omega_p-\Omega_k-\Omega_{k'})}{\Omega_k\Omega_{k'}\omega_p}\nonumber\\
&&\times\left[n_p(1-N_{k'})-(1+n_p)N_{k'}\right]\label{alphapsi}.
\end{eqnarray}
Here, $m$ stands for the fermion mass $m(t)$, $n_p(t)$ is (approximately--see below)
the number density of scalars, $\bm{p}=\bm{k}+\bm{k}'$ is the momentum of a scalar
produced, say, by the collision of two fermions of momenta $\bm{k}$ and $\bm{k}'$,
and $\omega_p=\sqrt{\vert\bm{p}\vert^2+M^2}$. In this approximation, the first two
terms in equation (\ref{dndtfermion}) combine to give
\begin{eqnarray}\label{Scatpsi}
S_k^{\psi}(t)&=&-\Gamma_k(t)\left[2N_k(t)-1\right]+\alpha_k(t)\nonumber\\
&=&\frac{g_\chi^2}{32\pi^2}(M^2-4m^2)\int
d^3k'\frac{\delta(\omega_p-\Omega_k-\Omega_{k'})}{\Omega_k\Omega_{k'}\omega_p}\nonumber\\
&\times&\left[n_p(1-N_k)(1-N_{k'})-(1+n_p)N_kN_{k'}\right]
\end{eqnarray}
which we recognize as the Boltzmann scattering integral corresponding to the decay of
a scalar into two fermions and the inverse production process. The evolution of $n_p(t)$
is described by a similar Boltzmann-like equation
\begin{eqnarray}
\partial_tn_p&=&S^{\chi}_p(t)\nonumber\\
&=&-\frac{g_\chi^2}{16\pi^2}(M^2-4m^2)\int
d^3k'\frac{\delta(\omega_p-\Omega_k-\Omega_{k'})}{\Omega_k\Omega_{k'}\omega_p}\nonumber\\
&\times&\left[n_p(1-N_k)(1-N_{k'})-(1+n_p)N_kN_{k'}\right],\label{chievolution}
\end{eqnarray}
where now $\bm{k}=\bm{p}-\bm{k}'$.  This contains no particle creation term (analogous to the last term of (\ref{dndtfermion}))
owing to our assumption that $\chi(\bm{x},t)$ does not couple to the inflaton.

At this point, we have a closed system of evolution equations, consisting of (\ref{dndtfermion}),
(\ref{dnudtfermion}), (\ref{chievolution}) and the inflaton equation of motion, which we
now take to have the form
\begin{equation}
\ddot{\phi}+m_\phi^2\phi+g\langle\bar{\psi}\psi\rangle_{\mathrm{trunc}}=0.\label{eomne}
\end{equation}
Here, we have approximated $\langle\bar{\psi}(\bm{x},t)\psi(\bm{x},t)\rangle$ by its
lowest-order contribution, $-(2\pi)^{-3}\int d^3k\,\mathrm{Tr}H(t,t',\bm{k})$.  It consists
of a local function of $\phi(t)$, which contributes to the effective potential, together
with the term
\begin{equation}
\langle\bar{\psi}\psi\rangle_{\mathrm{trunc}}=4\int\frac{d^3k}{(2\pi)^3\Omega_k(t)}
\left[m(t)N_k(t)+k\nu_k^R(t)\right].\label{eomint}
\end{equation}
Since we are concerned with the frictional effects arising from
$\langle\bar{\psi}\psi\rangle_{\mathrm{trunc}}$, we simplify matters
by replacing the whole effective potential with $\frac{1}{2}m_\phi^2\phi^2$. For later use,
we observe that the total energy given by
\begin{equation}
E=\frac{1}{2}\dot{\phi}^2+\frac{1}{2}m_{\phi}^2\phi^2+\int
\frac{dkk^2}{2\pi^2}\left[\omega_k n_k+4\Omega_kN_k\right]\label{conservedenergy}
\end{equation}
is exactly conserved by this system of evolution equations.

In the interest of accuracy, two remarks are in order.  First, the representation
(\ref{propfourier}) of the full 2-point functions is valid for a CP-invariant theory.
In practice, this means only that CP-breaking interactions will not be resummed
by the counterterm $\mathcal{M}$.  Second, the quantities denoted in this section by
$N_k(t)$, $\nu_k(t)$ and $n_p(t)$ are functions that arise in the process of solving
the appropriate differential equations for spinor and scalar propagators.  Only in
the free-field limit can they be unambiguously identified in terms of expectation values
of creation and annihilation operators as in (\ref{densities}).
\subsection{Renormalization}\label{renorm}
The integral (\ref{eomint}) that appears in the non-equilibrium
equation of motion (\ref{eomne}) has a divergent contribution from
the term $\nu_k^R$ in the integrand.  To isolate this divergence,
consider the evolution equation (\ref{dnudtfermion}) in the limit that
$k$ is large.  Assuming that the thermal terms proportional to
$N_k$ and $\Gamma_k$ can be neglected (which is to be expected by
analogy with equilibrium field theory, see e.g. section 3.5 of Ref \cite{bellac},
but hard to prove here) we find
\begin{equation}
\partial_t\nu_k(t)\approx -2ik\nu_k(t)+\frac{g}{2k}\dot{\phi}.
\end{equation}
The solution can be written as
\begin{equation}
\nu_k(t)=e^{-2ikt}\left[\nu_k(0)+\frac{g}{2k}\int_0^te^{2ikt'}\dot{\phi}(t')dt'\right].
\end{equation}
and on repeatedly integrating by parts we find
\begin{eqnarray}
\nu_k(t)&=&e^{-2ikt}\left[\nu_k(0)+\frac{ig}{4k^2}\dot{\phi}(0)-\frac{g}{8k^3}\ddot{\phi}(0)\right]\nonumber\\
&-&\frac{ig}{4k^2}\dot{\phi}(t)+\frac{g}{8k^3}\ddot{\phi}(t)+\cal{O}(k^{-4}).
\label{nuexp}
\end{eqnarray}
Now, contributions to $\nu_k^R(t)$ that vanish no faster than $k^{-3}$ as $k\rightarrow\infty$
will yield a divergent integral in (\ref{eomint}).
To avoid this, we firstly choose initial conditions which have the
form
\begin{equation}
\nu_k(0)=-\frac{ig}{4k^2}\dot{\phi}(0)+\frac{g}{8k^3}\ddot{\phi}(0)+\cal{O}(k^{-4})
\end{equation}
for large $k$, so that the square bracket in (\ref{nuexp}) vanishes.
We are then left with the term
$(g/8k^3)\ddot{\phi}(t)$ which can be canceled by a wavefunction renormalization
$\phi\to Z^{1/2}\phi$ in the equation of motion (\ref{eomne}). This is, in fact, the
same wavefunction renormalization that would be required in a zero-temperature
quantum field theory with the Yukawa coupling (\ref{yukawa}).  With a suitable
choice for $Z$, the renormalized integral is
\begin{eqnarray}\label{renormpsi}
\langle\bar{\psi}\psi\rangle_{\mathrm{trunc}}^{\mathrm{ren}}
&=&4\int\frac{dkk^2}{2\pi^2}\left[\frac{m(t)N_k(t)+k\nu_k^R(t)}{\Omega_k(t)}\right.\nonumber\\
&&\left. -\frac{gk}{8(k^2+m_0^2)^2}\ddot{\phi}\right],
\end{eqnarray}
where $m_0$ is a constant. The corresponding renormalized expression
for the conserved total energy is
\begin{eqnarray}\label{renormenergy}
E^{\mathrm{ren}}&=&\frac{1}{2}\dot{\phi}^2+\frac{1}{2}m_{\phi}^2\phi^2+\int
\frac{dkk^2}{2\pi^2}\bigg[\omega_kn_k+4\Omega_kN_k\nonumber\\
&&-\frac{g^2k\dot{\phi}^2}{16(k^2+m_0^2)^2}\bigg].
\end{eqnarray}
Note that, since $\nu_k^R$ has a contribution of order $k^{-3}$, the
spectrum of particles created by the last term in (\ref{dndtfermion})
has a corresponding contribution of order $k^{-4}$, and so therefore
does $N_k$.  This gives a convergent integral in (\ref{renormpsi}),
but a divergent contribution to $E^{\mathrm{ren}}$, which is cancelled
by the last term in (\ref{renormenergy}).
\subsection{Adiabatic Approximation}\label{adiabatic}
With the approximations developed above, the functions $N_k(t)$
and $\nu_k(t)$ that appear in $\langle\bar{\psi}\psi\rangle_{\mathrm{trunc}}$
obey local evolution equations (\ref{dndtfermion}) and (\ref{dnudtfermion}),
but the solution to these equations is non-local, i.e.
$\langle\bar{\psi}\psi\rangle_{\mathrm{trunc}}$ evaluated at time $t$
depends on $\phi(t')$ at all times $t'$ prior to $t$. One can attempt to derive
a local approximation of the form
\begin{equation}
g\langle\bar{\psi}\psi\rangle_{\mathrm{trunc}}\approx \Delta V'(\phi)+\eta(\phi)\dot\phi,
\label{tdexp}
\end{equation}
which depends only on the values of $\phi$ and $\dot\phi$ at time $t$, by means
of a time-derivative expansion, but we show below that, as in the case of
an inflaton that decays into scalar particles \cite{lawrie02}, this does not work.

To facilitate a time-derivative expansion, we introduce a formal expansion
parameter $\epsilon$ multiplying time derivatives
\begin{eqnarray}
\epsilon\partial_tN_k(t)&=&S_k^{\psi}+\epsilon\frac{gk\partial_t\phi(t)}{\Omega_k^2(t)}\nu_k^R(t)\label{dndtfermions3}\\
\epsilon\partial_tn_p(t)&=&S_p^{\chi}\label{dnchidtfermions3}\\
\epsilon\partial_t\nu_k(t)&=&-2[i\Omega_k(t)+\Gamma_k(t)]\nu_k(t)\nonumber\\
&&-\epsilon\frac{gk\partial_t\phi(t)}{2\Omega_k^2(t)}\left[2N_k(t)-1\right],\label{nandnu}
\end{eqnarray}
where we have also used $\dot{m}(t)=g\partial_t\phi(t)$.
We now expand $N_k(t)$, $n_p(t)$ and $\nu_k(t)$ in powers of
$\epsilon$ around the equilibrium distributions $N_k^{\mathrm{eq}}(t)$,
$n_p^{\mathrm{eq}}(t)$ and $\nu_k^{\mathrm{eq}}(t)$
\begin{eqnarray}
N_k(t) &=&N_k^{\mathrm{eq}}(t)+\epsilon\delta N_k(t)+O(\epsilon^2)\\
n_p(t) &=&n_p^{\mathrm{eq}}(t)+\epsilon\delta n_p(t)+O(\epsilon^2)\\
\nu_k(t) &=&\nu_k^{\mathrm{eq}}(t)+\epsilon\delta
\nu_k(t)+O(\epsilon^2).\label{expansion}
\end{eqnarray}
Substituting this expansion into equations (\ref{dndtfermions3})-(\ref{nandnu}) we find
to leading order that $\nu_k^{\mathrm{eq}}=0$, while
\begin{equation}
N_k^{\mathrm{eq}}(t)=\left[e^{\beta\Omega_k(t)}+1\right]^{-1},\quad
n_p^{\mathrm{eq}}=\left[e^{\beta\omega_p}-1\right]^{-1}\label{eqdist}
\end{equation}
for some inverse temperature $\beta$ are the usual Fermi-Dirac and Bose-Einstein
distributions that make the Boltzmann scattering integrals vanish.  At
order $\epsilon$, $\delta N_k$ and $\delta n_p$ are solutions of the integral
equations
\begin{eqnarray}\label{dndtadiabatic}
\partial_tN_k^{\mathrm{eq}}&=&\int dk'\left.\frac{\delta S_k^{\psi}}{\delta N_{k'}}\right\vert_{\mathrm{eq}}\delta N_{k'}+\int dp'\left.\frac{\delta S_k^{\psi}}{\delta
n_{p'}}\right\vert_{\mathrm{eq}}\delta n_{p'}\nonumber\\
\partial_tn_p^{\mathrm{eq}}&=&\int dk'\left.\frac{\delta S_p^{\chi}}{\delta N_{k'}}\right\vert_{\mathrm{eq}}\delta N_{k'}+\int dp'\left.\frac{\delta S_p^{\chi}}{\delta
n_{p'}}\right\vert_{\mathrm{eq}}\delta n_{p'}\nonumber\\
\end{eqnarray}
where $\vert_{\mathrm{eq}}$ means that we set $N=N^{\mathrm{eq}}$ and $n=n^{\mathrm{eq}}$
after differentiation. However, these equations are not self-consistent and therefore have no
solution. It is easily seen from (\ref{Scatpsi}) and (\ref{chievolution}) that
\begin{equation}
S\equiv 2\int d^3k\, S_k^\psi+\int d^3p\, S_p^\chi=0.\label{sumrule}
\end{equation}
This reflects the fact that if $N=\int d^3k\, N_k$ is the total number density of fermions
of each helicity (and also the number of antifermions of each helicity) and
$n=\int d^3p\, n_p$ is the total number density of scalars, then the quantity $2N+n$ is
conserved by the processes $\chi\leftrightarrow\psi\bar\psi$. (For example, one fermion
of a given helicity is produced, on average, in every two $\chi$ decays.) In the presence
of particle production, this conservation law no longer holds, but the scattering integrals
still obey the sum rule (\ref{sumrule}). The integral equations (\ref{dndtadiabatic}) therefore
entail
\begin{eqnarray}
\hbox to 40pt{$\displaystyle 2\partial_t\int d^3k\,N_k^{\mathrm{eq}}+\partial_t\int d^3p\,n_p^{\mathrm{eq}}$}
&&\nonumber\\
&=&\int dk'\frac{\delta S}{\delta N_{k'}}\delta N_{k'}
+\int dp'\frac{\delta S}{\delta n_{p'}}\delta n_{p'}\nonumber\\
&=&0,
\end{eqnarray}
but this is not true for the equilibrium distributions (\ref{eqdist}).
Thus, although the evolution equations themselves are consistent (albeit
approximate), they do not admit a solution in the form of a time-derivative
expansion.  In principle, therefore, we cannot approximate
$\langle\bar\psi\psi\rangle_{\mathrm{trunc}}$ in the form (\ref{tdexp})
or the inflaton equation of motion in the local form (\ref{localeom}) by
means of a time-derivative expansion.

A local approximation to the equation of motion \textit{can}
however be obtained by resorting to a further approximation, which
consists in replacing $N_k(t)$ and $n_p(t)$ with their equilibrium values
(\ref{eqdist}) for the purpose of evaluating the functions
$\Gamma_k$, which we then denote by $\Gamma_k^{\mathrm{eq}}$, and $\alpha_k$.
When this is done, the $\mathrm{O}(\epsilon)$ equations for $\delta N_k$
and $\delta\nu_k$ are
\begin{eqnarray}
\partial_tN_k^{\mathrm{eq}}&=&-2\Gamma_k^{\mathrm{eq}}\delta N_k\\
0&=&-2[i\Omega_k+\Gamma_k^{\mathrm{eq}}]\delta\nu_k-\frac{gk\dot{\phi}}{2\Omega_k^2}
\times\left[2N_k^{\mathrm{eq}}-1\right].
\end{eqnarray}
On rearranging these, we find that the inflaton equation of motion can be written as
\begin{equation}
\ddot\phi+m_\phi^2\phi+\Delta V'(\phi)+\eta(\phi)\dot\phi\approx 0,\label{eomad}
\end{equation}
with
\begin{eqnarray}
\Delta
V'(\phi)&=&\int\frac{dkk^2}{2\pi^2\Omega_k}\left[m(t)N_k^{\mathrm{eq}}(t)+k\nu_k^{R,\mathrm{eq}}(t)\right]\nonumber\\
&=&\int\frac{dkk^2}{2\pi^2\Omega_k(t)}\left(\frac{m(t)}{e^{\beta\Omega_k(t)}+1}\right)\label{deltavad}
\end{eqnarray}
and
\begin{eqnarray}
\eta(\phi)&=&\dot{\phi}^{-1}\int\frac{dkk^2}{2\pi^2\Omega_k}\left[m(t)\delta
N_k(t)+k\delta\nu_k^R(t)\right]\nonumber\\
&=&\frac{g}{4\pi^2}\int\frac{dkk^2}{\Omega_k(t)^2}\bigg\{\frac{\beta
m(t)^2N_k^{\mathrm{eq}}(t)(1-N_k^{\mathrm{eq}}(t))}{\Gamma_k^{\mathrm{eq}}(t)}\nonumber\\
&&+\frac{\Gamma_k^{\mathrm{eq}}(t)}{[\Omega^2_k(t)
+\Gamma_k^{\mathrm{eq}2}(t)]}\frac{k^2}{2\Omega_k(t)}\left[1-2N_k^{\mathrm{eq}}(t)\right]\bigg\}.\label{etaad}
\end{eqnarray}
In the zero temperature limit ($N^{\mathrm{eq}}_k=0$) and taking
$\Gamma_k<<\Omega_k$ this expression agrees with that found in
\cite{bereraramos} (apart from a factor of $2$ which we have not
been able to account for). We refer to the approximation represented by
(\ref{eomad})-(\ref{etaad}), together with a suitable prescription for
determining the temperature $\beta^{-1}$, as the \textit{adiabatic approximation}.

To estimate time evolution over any extended period of time within the
adiabatic approximation, we must allow $\beta$ to change in an appropriate
manner. We obtain a suitable prescription by requiring that the energy
(\ref{conservedenergy}) be conserved by the adiabatic evolution when
$n_k$ and $N_k$ have their equilibrium values.
\section{Numerical investigation}\label{results}
The approximate nonequilibrium evolution equations developed in section
\ref{deriv} lead to the local equation of motion (\ref{eomad}) only
when we make the extra approximation of replacing self-energies
(from which the functions $\alpha_k$ and $\Gamma_k$ are derived) with
their values in a state of thermal equilibrium, together with the
time-derivative expansion which then becomes possible. On the face of it,
one might expect these further approximations to be fairly harmless, at least
for a system that does not evolve too fast, and does not stray too far from
thermal equilibrium.  However, the numerical calculations described in
\cite{lawrienumeric} for the case of an inflaton coupled to scalar particles
showed that the nonequilibrium evolution governed by a set of equations analogous
to those of section \ref{deriv} is not well represented by the corresponding
adiabatic approximation, even under circumstances when that approximation
might seem to be reasonably well justified.

We therefore discuss in this section a similar numerical comparison of the
time evolution generated by the nonequilibrium equations described in sections
\ref{deriv} and \ref{renorm}, with that generated by the adiabatic approximation
of section \ref{adiabatic}.  To discretize the nonequilibrium time evolution, we
use a semi-implicit method defined schematically by
\begin{equation}
\bm{x}_{n+1}=\bm{x}_n+\left[\left.\frac{d\bm{x}}{dt}\right\vert_n+\left.\frac{d\bm{x}}{dt}\right\vert_{n+1}\right]\frac{\delta
t}{2},
\end{equation}
where $\bm{x}$ represents the whole collection of variables
$\phi$, $\dot\phi$, $N_k$, $n_k$, and $\nu_k$ and $n$ labels time
steps.  In the absence of the $\bar\psi\psi\chi$ interaction, it
is straightforward to show that this method is stable for our
problem.  Crucially, it is also simple enough that we can implement
a discretized version of the renormalization scheme of section
\ref{renorm} that is exact up to rounding errors.  When
interactions are included, stability of the resulting set of
integro-differential equations is very hard to analyse.  We do in
fact see evidence of instability when the calculations are pursued for
sufficiently long times, but the results presented below are for
periods of time for which the effects of this instability appear to
be negligible.  Moreover, although the renormalized total energy
(\ref{renormenergy}) is exactly conserved in the continuum limit,
there is no local discretized version of this expression that is
exactly conserved by the discretized evolution equations.  As a
rough check on our computations, we have evaluated a discretized
version of (\ref{renormenergy}), but we do not expect it to be
exactly conserved.  In particular, divergent contributions to
the integral, which cancel by construction in the continuum limit,
are not guaranteed to cancel when time evolution is discretized.
In practice, all the initial conditions we have considered lead
to an oscillatory solution for $\phi(t)$; we find that energy
acquires an oscillatory component (with amplitude of order 5\%
of the total), consistent with incomplete cancellation of the
divergent integral proportional to $\dot\phi^2$, together with
a drift amounting to a few percent over the time intervals of
interest.

The adiabatic approximation is not affected by these issues. Here, no
renormalization is required;  semi-implicit discretization of the
one remaining differential equation, the equation of motion (\ref{eomad}),
leads to a stable problem; and exact conservation of energy is imposed
as the criterion for determining $\beta$.  Intuitively, this approximation
ought to be good when $\phi(t)$ is slowly varying and thermalization of
the created particles is efficient.  A rough criterion for efficient
thermalization is
\begin{equation}
\tau\left\vert dm/dt\right\vert\ll m
\end{equation}
or,
\begin{equation}
\sigma\equiv\left\vert g\dot{\phi}/m\Gamma_{k=0}\right\vert\ll 1
\label{thermcrit}
\end{equation}
where $\tau$ is a relaxation time, which we take to be $\tau=1/\Gamma_{k=0}$.

To obtain slow variation in $\phi(t)$, we would ideally like to
identify a set of parameters for which the motion is overdamped.
Within the adiabatic approximation, one way of achieving this
is to make the first term in the friction coefficient (\ref{etaad})
large, by making $\Gamma_k$ relatively small, but this tends
to make thermalization slow, at least according to the criterion
(\ref{thermcrit}).  The parameter space of couplings, masses and
initial conditions is, even for the simple model considered here,
too large to permit a systematic exploration, but we have not succeeded
in locating a set of parameter values that leads both to overdamped
motion and to a very small value of $\sigma$. (This was also true
of the investigation reported in \cite{lawrienumeric} for an inflaton
decaying into scalars; in the context of warm inflation, it seems
to be generally hard to devise simple field theory models that
exhibit efficient thermalization as well as other desirable
properties \cite{bereraramosgleiser,bereraramosgleiser2,bereraNP}.)

\begin{figure}
\includegraphics[width=0.9\linewidth]{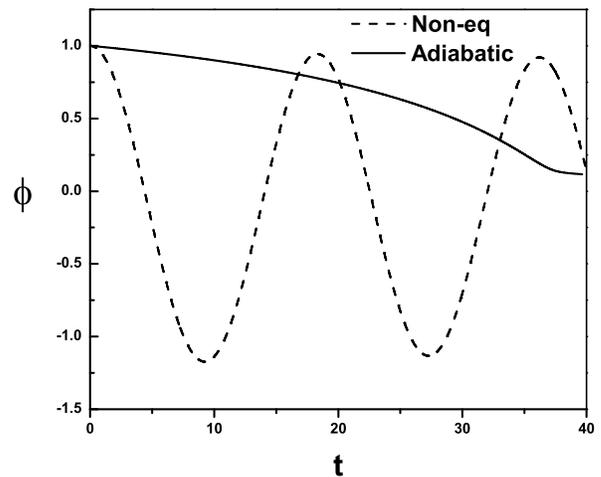}
\caption[Evolution of $\phi$ in both the adiabatic and
non-equilibrium solutions.] {Evolution of $\phi$ in both the
adiabatic and non-equilibrium solutions. The initial conditions
predict overdamping for the adiabatic solution.}\label{comparison}
\end{figure}

Nevertheless, we can find a situation in which the adiabatic approximation
predicts overdamped motion, and $\sigma$ is not too large.
Figure \ref{comparison} shows the evolution of $\phi(t)$
calculated from the nonequilibrium evolution equations and from
the adiabatic approximation, with the parameter values
$g=g_\chi=1$, $m_\phi=m_\chi=3$, $m_\psi=0.1$, and initial conditions
$\phi(0)=1$, $\dot\phi(0)=0$, $\beta=1$.  The solid curve is the
motion generated by the adiabatic approximation, which is indeed
overdamped;  we see, however, that the motion generated by the
non-equilibrium equations without the adiabatic approximation
(dashed curve) although it exhibits some damping is far from overdamped.

\begin{figure}
\includegraphics[width=0.9\linewidth]{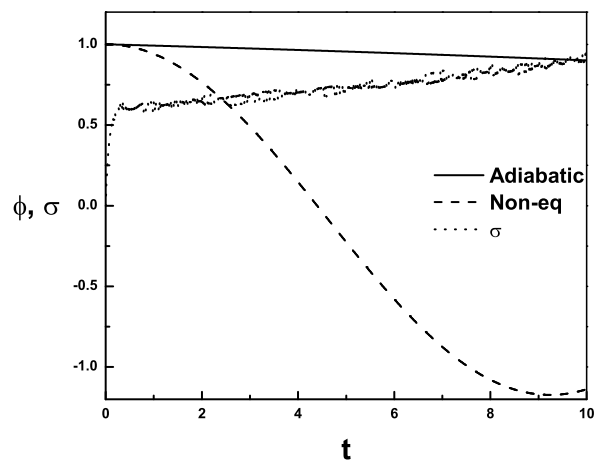}
\caption[Evolution of $\phi$ and $\sigma$ in the early stages.]
{Evolution of $\phi$ and $\sigma$ for early times. The
thermalization is relatively efficient in this period and the two
solutions are still very different}\label{comparisontherm}
\end{figure}
As in \cite{lawrienumeric}, this striking discrepancy does not
seem to be accounted for by inefficient thermalization.
In figure \ref{comparisontherm} we show the early stages of
the motion, along with the parameter $\sigma$ as estimated
within the adiabatic approximation. During this period of time,
$\sigma$ remains below 1.  Thus, within the adiabatic approximation,
one might well conclude that this approximation is reasonably
self-consistent.  While thermalization is not efficient enough
to guarantee good accuracy, one would probably not anticipate
the gross discrepancy between the two approximations for $\phi(t)$.
\begin{figure}
\includegraphics[width=0.9\linewidth]{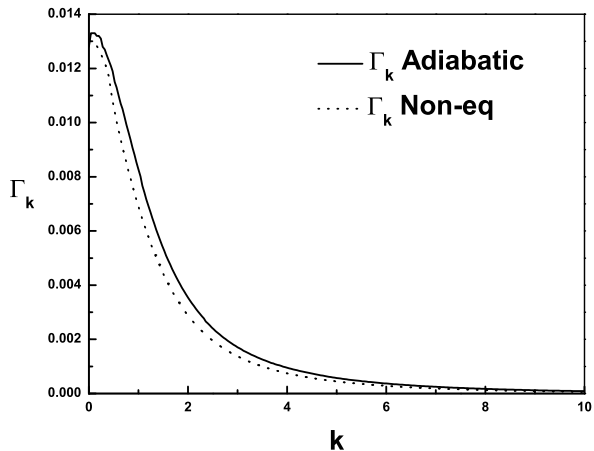}
\caption[Momentum distribution of $\Gamma_k$ at $t\approx 3$ for
both the adiabatic and the non-equilibrium solutions.] {Momentum
distribution of $\Gamma_k$ at $t\approx 3$ for both the adiabatic
and the non-equilibrium solutions.}\label{gammacomp}
\end{figure}
Moreover, we compare in figure \ref{gammacomp} the function $\Gamma_k$
calculated at $t\approx3$ from the nonequilibrium number densities
with that calculated in the adiabatic approximation from an exactly
thermal distribution.  While some difference is visible, it seems clear
that the small quantitative difference cannot in itself account for
difference between the overdamped and oscillatory motions
apparent in figure \ref{comparisontherm}.

In fact, the difference between these motions can be understood in
the way suggested in \cite{lawrienumeric}.  At $t=0$, we chose
an initial state of exact thermal equilibrium. Denote by $A(\phi)$
the integral (\ref{eomint}) evaluated in this state, and let
$U(\phi)=m_\phi^2\phi+A(\phi)$.  At early times,
when the value of $\phi$ has not changed by much, the nonequilibrium
equation of motion (\ref{eomne}) is given approximately by
\begin{equation}
\ddot\phi+U(\phi)\approx 0,\label{od1}
\end{equation}
while the adiabatic approximation to this equation is
\begin{equation}
\ddot\phi+U(\phi)+\eta(\phi)\dot\phi\approx 0.\label{od2}
\end{equation}
As seen from figure \ref{comparisontherm}, where
$\sigma\propto\vert\dot\phi\vert$, the velocity $\dot\phi$
in the adiabatic approximation quickly approaches the
``terminal velocity'' $\dot\phi\approx-U(\phi)/\eta(\phi)$
and thereafter $\phi$ and $\dot\phi$ evolve slowly, with
$\ddot\phi\approx 0$. According to the original equation, by
contrast, $\ddot\phi$ is still approximately equal to $-U(\phi)$,
leading to the dashed curve.  Clearly, in these circumstances,
the original equation of motion is not well represented by the
adiabatic approximation, and this is substantially independent
of how well the number densities used to calculate $U(\phi)$ and
$\eta(\phi)$ are thermalized.

\begin{figure}[htbp]
\includegraphics[width=0.9\linewidth]{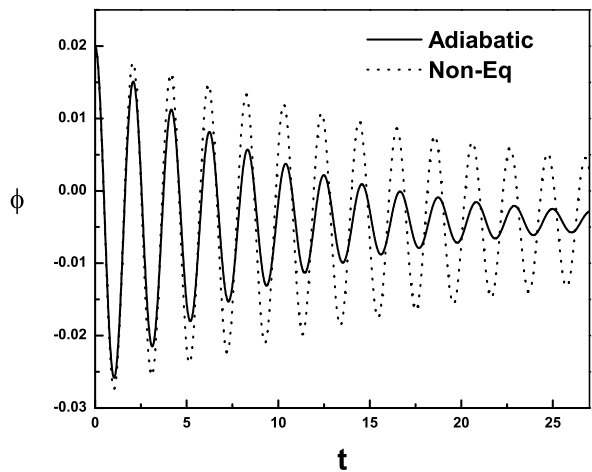}
\caption[Underdamped motion of $\phi$ for both the adiabatic and
non-equilibrium solutions] {Underdamped motion of $\phi$ for both
the adiabatic and non-equilibrium solutions}\label{underdamped}
\end{figure}

An example of the more generic case of underdamped motion is shown
in figure \ref{underdamped}.
We see that the nonequilibrium evolution shows significant damping,
but that the rate of damping is overestimated by the adiabatic
approximation.  This differs somewhat from the case of inflaton
decay into scalars \cite{lawrienumeric}.  In that case, the adiabatic
approximation leads to a friction coefficient roughly proportional
to $\phi^2$, and thus to an overestimate of damping at large amplitudes,
but an underestimate at small amplitudes.
\section{summary and discussion}\label{conclusions}
In this paper, we have sought to address a very specific issue:  can
the expectation value $\langle\bar\psi\psi\rangle$ appearing in the
generic inflaton equation of motion (\ref{phieom}), which is a non-local
functional of $\phi$, be adequately approximated by a local function
of $\phi(t)$ and $\dot\phi(t)$, leading to a local equation
of motion such as (\ref{localeom}).  The conclusion suggested by the
results of the two preceding sections is that it cannot be, but these results
are neither exact nor wholly rigorous, so it is worthwhile to summarize
carefully the line of argument we have adopted.

Derivations of the local friction coefficient $\eta(\phi)$ to be found
in the literature vary somewhat in the details of implementation, but
generally involve the {\it assumption}, characteristic of linear response
theory, that under suitable conditions the state of the particles can be
treated according to {\it equilibrium} field theory.  The possibility of
obtaining a local equation of motion then depends on the behaviour of the
relevant integral kernels, which can be investigated in perturbation theory
for specific models (see, for example \cite{bereraramos} and, for recent
discussions \cite{MossXiong,BereraMossRamos}).  We seek to avoid this
assumption by studying instead the slow-evolution limit of the
{\it nonequilibrium} state.  In principle, this is certainly the more
correct procedure;  in practice, however, nonequilibrium evolution is
extremely hard to deal with by analytic means.

The approximation scheme sketched in sections \ref{deriv} and \ref{renorm}
is the lowest non-trivial order of a partially resummed perturbation theory.
It leads to a local set of evolution equations, but we find that these equations
do not admit a solution in the form of a time-derivative expansion, which
is needed to derive the local equation of motion (\ref{localeom}), and consequently
that this local equation of motion does not, in principle, exist.
We cannot altogether rule out the possibility that this result is a feature
of our approximation scheme, rather than of the exact dynamics. Quantitatively,
indeed, the accuracy of the approximation is likely to be quite modest,
though we have no good way of assessing this.  However, the fact that it
reproduces the exact dynamics of section \ref{freefer} in the free-field limit,
and that the Boltzmann-like scattering terms have a sensible kinetic-theory
interpretation, offers some reassurance that the approximation captures
essential features of the nonequilibrium dynamics.  Moreover, any
improved method of studying nonequilibrium dynamics (for example, the
2-particle-irreducible formalism developed by Berges and coworkers;
see \cite{berges2} and references therein, especially \cite{berges1})
is intrinsically non-local. It seems very likely that any attempt to
derive a local equation of motion must follow a sequence of approximations
more or less equivalent to that used here, and would fail at the
same point.

Even if the local equation of motion is formally not valid, it might
nevertheless be a good approximation under suitable circumstances.
Indeed, we can recover essentially the same result for $\eta(\phi)$ as
that given by linear response methods by taking the further step of
replacing nonequilibrium self-energies with those calculated in a
state of thermal equilibrium:  we called this the adiabatic approximation
to our non-equilibrium evolution equations. Intuitively, this would seem
to be a fairly innocuous step if the evolution is fairly slow, and
thermalization is not too inefficient.  To test this, we obtained numerically,
in section \ref{results}, solutions for the motion of $\phi$ both with
and without the adiabatic approximation.  In order to focus on the frictional
effect (and to simplify the numerical calculation somewhat) we took the
whole effective potential (including the local part of $\langle\bar\psi\psi\rangle$)
to be approximated by the parabola $V=\frac{1}{2}m_\phi^2\phi^2$.  While
this might significantly affect the actual motion obtained for $\phi(t)$,
it is irrelevant for our purpose of assessing the effect of the extra adiabatic
approximation.

The results of this numerical comparison of the nonequilibrium and
adiabatic approximations are broadly similar to those of the parallel
calculations reported in \cite{lawrienumeric} for an inflaton decaying
into scalars.  In both cases, we find examples of conditions for which
the adiabatic approximation predicts overdamped motion (and so might seem,
at least on the grounds of self-consistency, to be reliable), while the motion
resulting from the ``full'' nonequilibrium evolution is oscillatory, with
relatively little damping.  (In neither case did we find conditions under
which the frictional effect generated by the nonequilibrium evolution is
sufficient to cause overdamping; however, we were unable to explore the
large parameter space exhaustively.)  Even with the adiabatic approximation,
underdamped motion seems to be more typical of the model we studied, but
the magnitude of the frictional effect seems to be significantly overestimated
by this approximation.  For the model studied in this work, indeed, it seems
that friction is always overestimated by the adiabatic approximation.
In the case of inflaton decay into scalars, the adiabatic approximation
yields a friction coefficient that contains a factor $\phi^2$, and appears
to overestimate the frictional effect when $\phi$ is large, while underestimating
it when $\phi$ is small.

Beyond the formal difficulty discussed in section \ref{adiabatic}, the reasons
for the failure of the adiabatic approximations are hard to identify in
any detail.  Poor thermalization may be an issue, but as noted earlier, it
seems implausible that the large discrepancies we observe are due
merely to the quantitative effect of number densities that differ somewhat
from a true thermal distribution.  A more plausible possibility is that
the discrepancies could be lessened simply by retaining more terms in the time-derivative
expansion. Comparison of (\ref{od1}) and (\ref{od2}) makes it rather obvious that,
if the adiabatic motion is friction limited, so that $\dot\phi\approx-U(\phi)/\eta(\phi)$
and $\ddot\phi\approx 0$, and if the same value of $\phi$ and the same number densities
are used to evaluate both the adiabatic and the ``full'' equation of motion,
a large discrepancy is more or less inevitable.  It could be, however, that
this friction-limited motion does not lead to negligible values for higher-order
terms in the time-derivative expansion, and that this accounts for the discrepancy.
Unfortunately, pursuing the adiabatic approximation to higher orders is rather
complicated (and not entirely unambiguous, in view of the evolving temperature
we employed); and to our knowledge, approximation schemes of this kind are not
commonly used in the cosmological literature.

The possibility of describing the dissipation of inflaton energy by a local
friction term is primarily of interest in connection with warm inflation
scenarios, and it should be emphasized that the results reported here and in
\cite{lawrienumeric} do not in themselves determine whether or not such
scenarios may be viable.  To the extent that any generic result can be extracted,
it seems that the adiabatic approximation overestimates the effects of
dissipation: in particular, overdamped motion (and hence warm inflation) is
apparently harder to achieve than the adiabatic approximation suggests.
However, this may be true only of the simple models we have investigated.
The somewhat discouraging conclusion is that no simple analytical approximation
known to us represents dissipation reliably, and numerical investigations
of the kind reported here offer a very inefficient tool for inflationary
model building.


\begin{thebibliography}{23}
\expandafter\ifx\csname natexlab\endcsname\relax\def\natexlab#1{#1}\fi
\expandafter\ifx\csname bibnamefont\endcsname\relax
  \def\bibnamefont#1{#1}\fi
\expandafter\ifx\csname bibfnamefont\endcsname\relax
  \def\bibfnamefont#1{#1}\fi
\expandafter\ifx\csname citenamefont\endcsname\relax
  \def\citenamefont#1{#1}\fi
\expandafter\ifx\csname url\endcsname\relax
  \def\url#1{\texttt{#1}}\fi
\expandafter\ifx\csname urlprefix\endcsname\relax\def\urlprefix{URL }\fi
\providecommand{\bibinfo}[2]{#2}
\providecommand{\eprint}[2][]{\url{#2}}

\bibitem[{\citenamefont{Moss}(1985)}]{moss2}
\bibinfo{author}{\bibfnamefont{I.~G.} \bibnamefont{Moss}},
  \bibinfo{journal}{Phys. Lett.} \textbf{\bibinfo{volume}{154B}},
  \bibinfo{pages}{120} (\bibinfo{year}{1985}).

\bibitem[{\citenamefont{Berera}(1995)}]{berera}
\bibinfo{author}{\bibfnamefont{A.}~\bibnamefont{Berera}},
  \bibinfo{journal}{Phys. Rev. Lett.} \textbf{\bibinfo{volume}{75}},
  \bibinfo{pages}{3218} (\bibinfo{year}{1995}).

\bibitem[{\citenamefont{Berera and Fang}(1995)}]{bererafang}
\bibinfo{author}{\bibfnamefont{A.}~\bibnamefont{Berera}} \bibnamefont{and}
  \bibinfo{author}{\bibfnamefont{L.}~\bibnamefont{Fang}},
  \bibinfo{journal}{Phys. Rev. Lett.} \textbf{\bibinfo{volume}{74}},
  \bibinfo{pages}{1912} (\bibinfo{year}{1995}).

\bibitem[{\citenamefont{Berera}(2000)}]{bereraNP}
\bibinfo{author}{\bibfnamefont{A.}~\bibnamefont{Berera}},
  \bibinfo{journal}{Nucl. Phys. B} \textbf{\bibinfo{volume}{585}},
  \bibinfo{pages}{666} (\bibinfo{year}{2000}).

\bibitem[{\citenamefont{Hall et~al.}(2004)\citenamefont{Hall, Moss, and
  Berera}}]{moss}
\bibinfo{author}{\bibfnamefont{L.}~\bibnamefont{Hall}},
  \bibinfo{author}{\bibfnamefont{I.~G.} \bibnamefont{Moss}}, \bibnamefont{and}
  \bibinfo{author}{\bibfnamefont{A.}~\bibnamefont{Berera}},
  \bibinfo{journal}{Phys. Rev. D} \textbf{\bibinfo{volume}{69}},
  \bibinfo{pages}{083525} (\bibinfo{year}{2004}).

\bibitem[{\citenamefont{Hosoya and Sakagami}(1984)}]{hosoya}
\bibinfo{author}{\bibfnamefont{A.}~\bibnamefont{Hosoya}} \bibnamefont{and}
  \bibinfo{author}{\bibfnamefont{M.}~\bibnamefont{Sakagami}},
  \bibinfo{journal}{Phys. Rev. D} \textbf{\bibinfo{volume}{29}},
  \bibinfo{pages}{2228} (\bibinfo{year}{1984}).

\bibitem[{\citenamefont{Morikawa and Sasaki}(1984)}]{morikawa}
\bibinfo{author}{\bibfnamefont{M.}~\bibnamefont{Morikawa}} \bibnamefont{and}
  \bibinfo{author}{\bibfnamefont{M.}~\bibnamefont{Sasaki}},
  \bibinfo{journal}{Prog. Theor. Phys.} \textbf{\bibinfo{volume}{72}},
  \bibinfo{pages}{782} (\bibinfo{year}{1984}).

\bibitem[{\citenamefont{Morikawa and Sasaki}(1985)}]{morikawa2}
\bibinfo{author}{\bibfnamefont{M.}~\bibnamefont{Morikawa}} \bibnamefont{and}
  \bibinfo{author}{\bibfnamefont{M.}~\bibnamefont{Sasaki}},
  \bibinfo{journal}{Phys. Lett. B} \textbf{\bibinfo{volume}{165}},
  \bibinfo{pages}{59} (\bibinfo{year}{1985}).

\bibitem[{\citenamefont{Berera et~al.}(1998)\citenamefont{Berera, Ramos, and
  Gleiser}}]{bereraramosgleiser}
\bibinfo{author}{\bibfnamefont{A.}~\bibnamefont{Berera}},
  \bibinfo{author}{\bibfnamefont{R.}~\bibnamefont{Ramos}}, \bibnamefont{and}
  \bibinfo{author}{\bibfnamefont{M.}~\bibnamefont{Gleiser}},
  \bibinfo{journal}{Phys. Rev. D} \textbf{\bibinfo{volume}{58}},
  \bibinfo{pages}{123508} (\bibinfo{year}{1998}).

\bibitem[{\citenamefont{Berera and Ramos}(2001)}]{bereraramos}
\bibinfo{author}{\bibfnamefont{A.}~\bibnamefont{Berera}} \bibnamefont{and}
  \bibinfo{author}{\bibfnamefont{R.}~\bibnamefont{Ramos}},
  \bibinfo{journal}{Phys. Rev. D} \textbf{\bibinfo{volume}{63}},
  \bibinfo{pages}{103509} (\bibinfo{year}{2001}).

\bibitem[{\citenamefont{Lawrie}(2002)}]{lawrie02}
\bibinfo{author}{\bibfnamefont{I.~D.} \bibnamefont{Lawrie}},
  \bibinfo{journal}{Phys. Rev. D} \textbf{\bibinfo{volume}{66}},
  \bibinfo{pages}{041702(R)} (\bibinfo{year}{2002}).

\bibitem[{\citenamefont{Lawrie}(2003)}]{lawrie}
\bibinfo{author}{\bibfnamefont{I.~D.} \bibnamefont{Lawrie}},
  \bibinfo{journal}{Phys. Rev. D} \textbf{\bibinfo{volume}{67}},
  \bibinfo{pages}{045006} (\bibinfo{year}{2003}).

\bibitem[{\citenamefont{Lawrie}(2005)}]{lawrienumeric}
\bibinfo{author}{\bibfnamefont{I.~D.} \bibnamefont{Lawrie}},
  \bibinfo{journal}{Phys. Rev. D} \textbf{\bibinfo{volume}{71}},
  \bibinfo{pages}{025021} (\bibinfo{year}{2005}).

\bibitem[{\citenamefont{Lawrie and McKernan}(2000)}]{lawriefermion}
\bibinfo{author}{\bibfnamefont{I.~D.} \bibnamefont{Lawrie}} \bibnamefont{and}
  \bibinfo{author}{\bibfnamefont{D.~B.} \bibnamefont{McKernan}},
  \bibinfo{journal}{Phys. Rev. D} \textbf{\bibinfo{volume}{62}},
  \bibinfo{pages}{105032} (\bibinfo{year}{2000}).

\bibitem[{\citenamefont{Semenoff and Weiss}(1985)}]{semenoff}
\bibinfo{author}{\bibfnamefont{G.~W.} \bibnamefont{Semenoff}} \bibnamefont{and}
  \bibinfo{author}{\bibfnamefont{N.}~\bibnamefont{Weiss}},
  \bibinfo{journal}{Phys. Rev. D} \textbf{\bibinfo{volume}{31}},
  \bibinfo{pages}{689} (\bibinfo{year}{1985}).

\bibitem[{\citenamefont{Kobes et~al.}(1985)\citenamefont{Kobes, Semenoff, and
  Weiss}}]{kobes}
\bibinfo{author}{\bibfnamefont{R.~L.} \bibnamefont{Kobes}},
  \bibinfo{author}{\bibfnamefont{G.~W.} \bibnamefont{Semenoff}},
  \bibnamefont{and} \bibinfo{author}{\bibfnamefont{N.}~\bibnamefont{Weiss}},
  \bibinfo{journal}{Z. Phys. C} \textbf{\bibinfo{volume}{29}},
  \bibinfo{pages}{371} (\bibinfo{year}{1985}).

\bibitem[{\citenamefont{Lawrie}(1989)}]{lawrie2}
\bibinfo{author}{\bibfnamefont{I.~D.} \bibnamefont{Lawrie}},
  \bibinfo{journal}{Phys. Rev. D} \textbf{\bibinfo{volume}{40}},
  \bibinfo{pages}{3330} (\bibinfo{year}{1989}).

\bibitem[{\citenamefont{le~Bellac}(1996)}]{bellac}
\bibinfo{author}{\bibfnamefont{M.}~\bibnamefont{le~Bellac}},
  \emph{\bibinfo{title}{Thermal Field Theory}} (\bibinfo{publisher}{Cambridge
  University Press}, \bibinfo{year}{1996}).

\bibitem[{\citenamefont{Berera et~al.}(1999)\citenamefont{Berera, Ramos, and
  Gleiser}}]{bereraramosgleiser2}
\bibinfo{author}{\bibfnamefont{A.}~\bibnamefont{Berera}},
  \bibinfo{author}{\bibfnamefont{R.}~\bibnamefont{Ramos}}, \bibnamefont{and}
  \bibinfo{author}{\bibfnamefont{M.}~\bibnamefont{Gleiser}},
  \bibinfo{journal}{Phys. Rev. Lett.} \textbf{\bibinfo{volume}{83}},
  \bibinfo{pages}{264} (\bibinfo{year}{1999}).

\bibitem[{\citenamefont{Moss and Xiong}()}]{MossXiong}
\bibinfo{author}{\bibfnamefont{I.~G.} \bibnamefont{Moss}} \bibnamefont{and}
  \bibinfo{author}{\bibfnamefont{C.}~\bibnamefont{Xiong}},
  \eprint{arXiv:hep-th/0603266}.

\bibitem[{\citenamefont{Berera et~al.}()\citenamefont{Berera, Moss, and
  Ramos}}]{BereraMossRamos}
\bibinfo{author}{\bibfnamefont{A.}~\bibnamefont{Berera}},
  \bibinfo{author}{\bibfnamefont{I.~G.} \bibnamefont{Moss}}, \bibnamefont{and}
  \bibinfo{author}{\bibfnamefont{R.~O.} \bibnamefont{Ramos}},
  \eprint{arXiv:0706.2793}.

\bibitem[{\citenamefont{Berges et~al.}(2003)\citenamefont{Berges, Bors\'anyi,
  and Serreau}}]{berges2}
\bibinfo{author}{\bibfnamefont{J.}~\bibnamefont{Berges}},
  \bibinfo{author}{\bibfnamefont{S.}~\bibnamefont{Bors\'anyi}},
  \bibnamefont{and} \bibinfo{author}{\bibfnamefont{J.}~\bibnamefont{Serreau}},
  \bibinfo{journal}{Nucl. Phys B} \textbf{\bibinfo{volume}{660}},
  \bibinfo{pages}{51} (\bibinfo{year}{2003}).

\bibitem[{\citenamefont{Berges and Bors\'anyi}(2007)}]{berges1}
\bibinfo{author}{\bibfnamefont{J.}~\bibnamefont{Berges}} \bibnamefont{and}
  \bibinfo{author}{\bibfnamefont{S.}~\bibnamefont{Bors\'anyi}},
  \bibinfo{journal}{Nucl. Phys A} \textbf{\bibinfo{volume}{785}},
  \bibinfo{pages}{58} (\bibinfo{year}{2007}).

\end{thebibliography}

\end{document}